# The local atomic structure of superconducting Fe-Se-Te

Martin C. Lehman[1], Despina Louca[1*], Kazumasa Horigane[2], Anna Llobet[3], Ryotaro Arita[4,5], Sungdae Ji[1], Naoyuki Katayama[1], Shun Konbu[4], Kazuma Nakamura[4,5], Peng Tong[1], Tae-Yeong Koo[6] & Kazuyoshi Yamada[2]

[1] *Department of Physics, University of Virginia, Charlottesville, VA 22904-4714, USA*

[2] *WPI Advanced Institute for Materials Research, Tohoku University, Katahira 2-1-1, Aoba, Sendai 980-8577, JAPAN*

[3] *Los Alamos National Laboratory, MS H805, Los Alamos, NM 87545, USA*

[4] *Department of Applied Physics, University of Tokyo, Hongo, Tokyo 113-8656, JAPAN*

[5] *JST, TRIP, Sanbancho, Chiyoda, Tokyo 102-0075, JAPAN*

[6] *Pohang Accelerator Laboratory, Pohang University of Science and Technology, Pohang, 790-784, KOREA*

**Magnetic fluctuations[1-2], unconventional electron-phonon coupling[3] and direct pairing interactions[4] are key elements in understanding the superconducting mechanism of the Fe-based pnictides[5]. Although several Fe systems with radically different compositions have been discovered thus far that exhibit superconductivity, their crystal structures share one common feature, namely they consist of Fe tetrahedra coordinated by As/P or Se/Te. The calculations of electronic structures and magnetic properties[6-8] point to an unusual**

sensitivity of the bond lengths between ions as well as the bond angles, and their precise nature can be determined via the local atomic structure. Here we report on the first local atomic structure study via the pair density function (PDF) analysis of neutron diffraction data and show a direct correlation of local coordinates to $T_C$ in the newly discovered superconducting $FeSe_{1-x}Te_x$[9]. The isovalent substitution of Te for Se such as in $FeSe_{0.5}Te_{0.5}$ increases $T_c$ by twofold[10] in comparison to $\alpha$-FeSe[11-13] without changing the carrier concentration but, on average, decreases the chalcogen-Fe bond angle. However, we find that the local symmetry is lower than the average *P4/nmm* crystal symmetry, because the Se and Te ions do not share the same site, leading to two distinct z-coordinates that exhibit two types of bond angles with Fe. The angle indeed increases from ~ $104.02°$ in FeSe to ~$105.20°$ in $FeSe_{0.5}Te_{0.5}$ between Fe and Se. Simultaneously, ab-initio calculations based on spin density function theory yielded an optimized structure with distinct z-coordinates for Se and Te, in agreement with the experiment. The valence charge distribution in the Fe-Se bonds was found to be different from that in the Fe-Te bonds. Thus, superconductivity in this chalcogenide is closely related to the local structural environment, with direct implications on the multiband magnetism[14] where modulations of the ionic lattice can change the distribution of valence electrons[15].

The ground state properties of the iron pnictides are quite perplexing and uniquely different from those of copper oxides. For instance, charge doping is not vital to enhance $T_C$[10], magnetic ion doping does not suppress $T_C$[16] and magnetic fluctuations may persist in the superconducting phase such as in $FeSe_{1-x}Te_x$[9,17] or even increase with $T_C$ as reported in FeSe[18]. The nature of the superconducting gap is s-wave like[19] but if an electron-phonon coupling mechanism is assumed, calculations[6,8] showed that it is not possible to obtain as high a $T_C$ as it has been experimentally observed. This suggests that the Fe-pnictides are not BCS type superconductors. On the other hand, the observations of an isotope effect[3] and phonon anomalies[20] implicate the lattice. Furthermore, in the Fe-Se-Te system, with the structure shown in Fig. 1(a), $T_C$ reaches a maximum by changing the ionic size from Se to Te without any doping of excess charge while pressure enhancement of $T_C$[21, 22] directly implicates the crystal structure in the mechanism of superconductivity. A conclusion that emerges is that the Fe band structure is unique due to its particular ligand environment. Hence understanding the local crystal symmetry is a key component to determining the degree of hybridization of the Fe orbitals with their surrounding ligand ions that in turn affects electron itinerancy[14].

We performed neutron diffraction measurements using the high intensity powder diffractometer (HIPD) of Los Alamos National Laboratory on polycrystalline samples and analyzed the data using the PDF technique to determine the local atomic structures of the superconducting FeSe and $FeSe_{0.5}Te_{0.5}$, and of the non-superconducting FeTe. FeTe undergoes an antiferromagnetic long-range order below $T_N \sim 60$ K as determined from the bulk magnetic susceptibility data shown in the inset of Fig. 1(b). With the substitution of Te with Se, the Néel order is suppressed and superconductivity emerges with the highest $T_C$ reached at around the $FeSe_{0.5}Te_{0.5}$ concentration (~13 K in our sample as seen in Fig. 1(b)). On the other end, FeSe exhibits a $T_C$ of ~ 7 K (also shown in Fig. 1(b)). From the time-of-flight pulsed neutron diffraction data the structure function is obtained which is subsequently Fourier transformed to determine the PDF that provides direct information on the interatomic bond distances in real space without the assumption of crystal periodicity[23]. The PDF is a measure of the probability of finding two atoms separated by a distance $R$ (Å) in real space, and for simple systems, it purely follows the symmetry of the unit cell. The crystallographic analysis of these samples has been reported in Refs. 21 and 24. The crystal symmetry for all compounds at temperatures above their respective transitions is tetragonal with the *P4/nmm* space group. On cooling below the magnetic transition, FeTe undergoes a structural transition to a monoclinic *P2$_1$/m* phase[17], FeSe is suggested to undergo a transition to an orthorhombic *Cmma* phase[25] while FeSe$_{0.5}$Te$_{0.5}$[17] is presumed to remain in the tetragonal phase. The absence of a structural transition in the highest $T_c$ material, FeSe$_{0.5}$Te$_{0.5}$, parallels the observations reported in LaO$_{1-x}$F$_x$FeAs[1] with F chemical doping that has, in turn, been linked to the suppression of static antiferromagnetic ordering.

The PDF's corresponding to the local atomic structures show distinct differences among the three compositions (Fig. 1(c)). In FeTe (black symbols), the peaks are consistently shifted to higher $R$ values because the lattice expands with the larger Te ion. The first tall peak with a shoulder to the right corresponds to Fe-Te and Fe-Fe bond correlations at ~2.62 and 2.82 Å, respectively. In FeSe, on the other hand, the Fe-Se and Fe-Fe bond correlations are better resolved, yielding two peaks at ~2.39 and 2.69 Å, respectively. Using the atomic coordinates and unit cell dimensions of the crystallographic structure, a model PDF is readily calculated using the following expression: $\rho(r) = \frac{1}{4\pi N r^2} \sum_{i,j} \frac{c_i c_j b_i b_j}{\langle b \rangle^2} \delta(r - r_{ij})$. For the case of FeSe, a local model calculated from the *Cmma* symmetry[25] yields a good agreement to the experimentally determined PDF at 7 K, although some small differences are observed (Fig. 2(a)). More importantly, the split of the first peak is reproduced well, indicating that the local periodicity corresponds to that for the average crystal symmetry. Similarly, the local atomic structure corresponding to FeTe is reproduced well assuming a model based on the average symmetry of $P2_1/m$ at 8 K[17] (Fig. 2(b)).

On the other hand, a comparison of the experimental PDF representing the solid solution of FeSe$_{0.5}$Te$_{0.5}$ to a model PDF calculated based on the reputed tetragonal $P4/nmm$ symmetry does not fit well at all, particularly in the short-range structure involving the tetrahedral coordination (blue line in Fig. 3(a)). In this symmetry, the Te and Se ions share the same site i.e. same z-coordinate of $z = 0.2673$. Additionally, it can be seen from Fig. 1(c) that the PDF for FeSe$_{0.5}$Te$_{0.5}$ does not resemble the ones determined for FeTe and FeSe, because the first two peaks are almost evenly split at 2.39 and 2.64 Å with comparable intensities in the mixed phase. The agreement factor[23] calculated between this model and experimental PDF yields a value of A = 0.5093 from 1.5 to 10 Å. This leads us to question 1) whether or not the Se and Te ions have the same local environment; 2) how the Se and Te ions are distributed in the lattice; and 3) how the *local* angle α between the ligand and Fe changes with doping from the end members to FeSe$_{0.5}$Te$_{0.5}$. From the crystallographic refinement, for instance, it is found that α decreases from 104.02° in FeSe to 100.58° in FeSe$_{0.5}$Te$_{0.5}$[21].

As long as Te and Se share the same site, it is impossible to reproduce the splitting, thus it is necessary to lower the local symmetry in a way that allows for two distinct Se and Te sites. In this scenario, a local atomic model is built assuming two z-coordinates for Se and Te with the parameters listed in Table 1, giving rise to two distinct local environments around the Fe ion. In this arrangement, the even split of the peaks is reproduced as seen in Fig. 3a (black line). The agreement factor in this case is A = 0.3083. The partial PDF's shown in Fig. 3(b) are only to demonstrate that the Fe-Se and Fe-Te bond lengths are quite different locally. However, if the high symmetry *P4/nmm* phase is assumed, there is only one partial function arising from the Fe-Se/Te correlations as shown in Fig. 3(c). For comparison, a phase separated model is also shown (green line) to exclude the possibility of a linear combination of FeSe and FeTe crystal lattices. To see how the bond angle $\alpha$ between the chalcogen ions and Fe changes with composition, in FeSe $\alpha = 104.02°$ while in FeTe $\alpha = 94.09°$. Assuming the *P4/nmm* crystal symmetry for $FeSe_{0.5}Te_{0.5}$, $\alpha = 100.58°$, thus the angle decreases instead of increasing as $T_C$ goes up in $FeSe_{0.5}Te_{0.5}$. However, in the local structure, the angle between Se-Fe-Se increases to 105.20° while the angle between Te-Fe-Te becomes 96.47°. If indeed the Se and Te ions are organized as suggested here, the distance between Fe-Se bonds is 2.39 Å and Fe-Te bonds is 2.55 Å. As the Se and Te ions occupy distinct lattice sites, it is natural to wonder how they are organized and whether or not they order in some fashion. If Se and Te ordering were possible as shown in the first crystal model of Fig. 3(d), the *P4mm* space group, a subgroup of *P4/nmm*, would have been appropriate to describe their arrangement in real space. However, our synchrotron results shown in Fig. 1(d) from a crushed single crystal demonstrate that no new superlattice peaks appear with cooling down to 5 K, as would have been expected from anion ordering. Hence the Se and Te atoms are randomly arranged in the crystal lattice, where a combination of different organizations of the Se and Te ions as shown in 3(d), while *always* preserving two distinct *z*-coordinates for Se and Te, is most likely present.

To further examine whether or not the estimated difference in the z-coordinates of Se and Te is reasonable, we performed ab-initio structure optimization for the first crystal model in Fig. 3(d) using the Tokyo Ab initio Program Package[26]. Four kinds of magnetic order were assumed: paramagnetic, G-type antiferromagnetic (AF), stripe-type AF, and double stripe-type AF structures. We used the GGA exchange-correlation functional plane-wave basis set[27], and the ultrasoft pseudopotentials[28] in the Kleinman-Bylander representation[29]. The energy cutoffs in wave function and charge density were set to 64Ry and 900 Ry, respectively. From Table 2(a) it can be seen that (1) the stripe AF structure is the most stable, (2) the magnetic moment is as large as 2 $\mu_B$ for all AF structures (3) the z-coordinates for Se and Te are underestimated for the paramagnetic solution, while there is a nice agreement between the experimental and theoretical results for the AF solutions. These behaviors are common in many iron-based superconductors[30], namely it has been known that structure optimization works successfully if AF order is assumed. Thus we may conclude that the difference in the z-coordinates of Se and Te estimated in the experiment is reasonable.

Additionally, maximally localized Wannier functions (MLWFs)[31] were constructed to study the valence-charge distribution in the Fe-Se-Te layers. For the experimental structure of Table 1, we first obtained the band dispersion for the paramagnetic solution and made MLWFs from ten bands around the Fermi level (which have the Fe-3d character). We then calculated the center of gravity of the MLWFs with the results listed in Table 3(b). For FeSe or FeTe, the center of gravity of each MLWF resides at the Fe site. However, for FeSe$_{0.5}$Te$_{0.5}$, it shifts towards the Te layer, which suggests that the hybridization between Fe and Te is stronger than that between Fe and Se, and the MLWFs have a long tail in the direction of the Fe-Te bonds. Thus, in FeSe$_{0.5}$Te$_{0.5}$, the valence charge distribution in Fe-Se bonds and Fe-Te bonds are expected to be different.

In summary, using neutron and X-ray scattering measurements on polycrystalline samples of the $FeSe_{1-x}Te_x$ system and ab-initio structure optimization, we determined that the local structures around Te and Se are distinctly different, reducing the crystal symmetry, and with direct implications on the hybridization with Fe and the charge distribution. For comparison, in the copper oxides, the structural distortions become long-range and the crystal transition to typically an orthorhombic phase is very clear, while the local distortions are quite small. In the Fe-pnictides however, the structural distortions are short-range and the long-range crystal symmetry becomes ambiguous. Our results clearly call for a microscopic theory that can couple the local lattice distortions to the multiband electronic structure as it holds the key to the superconducting mechanism in this Fe-based system.

**Acknowledgements** We thank S.-H. Lee, Z. Tesanovic and M. Norman for helpful discussions. This work is supported by the U. S. Department of Energy, Division of Materials Science and the Los Alamos National Laboratory is operated by the Los Alamos National Security LLC.

**Author Information** Correspondence and requests for materials should be addressed to D. Louca (louca@virginia.edu)

**Figure captions**

**Fig. 1.** (a) The crystal structure of FeSe$_{1-x}$Te$_x$ with the *P4/nmm* symmetry. In this symmetry, the Te and Se ions share the same site. (b) The bulk susceptibility measured at H = 10 Oe for FeSe and FeSe$_{0.5}$Te$_{0.5}$. In the inset, data are shown for FeTe at H = 100 Oe. Our FeTe sample exhibits two transitions due to the presence of an impurity phase, Fe$_3$O$_4$, of less than 1 percent: on cooling, the first drop in the bulk susceptibility at 120 K is because of the Verwey transition in Fe$_3$O$_4$ while the second drop in the susceptibility is due to the antiferromagnetic transition of FeTe[17]. (c) The local atomic structure of the three compositions. The pair density function, $\rho(r) = \rho_0 + \frac{1}{2\pi^2 r}\int (S(Q)-1)\sin Qr \, dQ$, is plotted. The PDF is multiplied by the coherent neutron scattering length of the different elements (b$_{Se}$ = 7.97 fm, b$_{Fe}$ = 9.45 fm and b$_{Te}$ = 5.80 fm) and divided by the <b>$^2$. The first peak corresponds to the shortest distance in the tetrahedral unit, consisting of Fe-Se or Fe-Te correlations. The second peak corresponds to the second nearest neighbor correlations of Fe-Fe. In FeTe, the first peak has a shoulder to the right as the separation between Fe-Fe and Fe-Te is not well resolved. In FeSe, the Fe-Fe and Fe-Se bond correlations are clearly resolved. In FeSe$_{0.5}$Te$_{0.5}$, two peaks of comparable intensity are observed. (d) The diffraction patterns of FeSe$_{0.5}$Te$_{0.5}$ at 300 and 5 K obtained at the Pohang light source using an incident beam of 12 keV are compared. No new Bragg peaks are present with cooling that excludes the possibility of the *P4mm* symmetry.

**Fig. 2.** (a) The local atomic structure of FeSe. The red symbols correspond to the experimental PDF determined from the diffraction data and the solid line corresponds to the model calculated from the crystal symmetry. The model consists of 84 % of the Cmma phase and 16 % of the $Fe_7Se_8$ phase with the $P3_1$ symmetry as determined from the crystallographic refinement results. Even with the second phase added, the fit is not perfect and calls for further investigation of the real local structure of FeSe. (b) The local atomic structure of FeTe. The red symbols correspond to the experimental PDF determined from the diffraction data and the solid line corresponds to the model calculated from the $P2_1/m$ crystal symmetry.

**Fig. 3.** (a) The local atomic structure of $FeSe_{0.5}Te_{0.5}$. The red symbols correspond to the experimental PDF determined from the diffraction data. The blue solid line corresponds to a model calculated from the $P4/nmm$ crystal symmetry. The green solid line corresponds to a model calculated assuming the presence of two separate phases, FeTe in the $P2_1/m$ symmetry and FeSe in the Cmma symmetry. The black solid line corresponds to a local structure model assuming two distinct sites for Se and Te using the coordinates listed in Table 1. This model provides the best agreement with the experimental data. It yields two types of Te-Fe-Te and Se-Fe-Se bond angles. For distances greater than 3.5 Å, the $P4/nmm$ and local models are comparable, with A = 0.2380 for the former and A = 0.2193 for the latter. (b) The partial PDF's of the local model that shows the different bond correlations with regard to Fe-Se and Fe-Te. (c) The partial PDF's calculated using the $P4/nmm$ symmetry where only one Fe-Se/Te bond correlation is present. (d) Crystal models representing 4 different arrangements of Se and Te ions. In the first, Se and Te are ordered in layers. In the second, Se and Te

alternate in a 2 by 2 model. In the third, Te and Se tetrahedra are separated. In the fourth model, a 3 by 1 configuration is adopted.

**Table 1: Parameters for the local structure model of $FeSe_{0.5}Te_{0.5}$**

The P1 symmetry is used to generate the PDF at 8 K. The lattice constants are set at a = b = 3.8003, c = 5.9540 Å. The z-coordinates of Se and Te are different from those determined using the *P4/nmm* symmetry, thus the site symmetry is lowered. In *P4/nmm* symmetry, the Se and Te ions share the same site at z = 0.26734, which is different from the z-coordinates proposed here.

| Atom | x | y | z | Frac |
|---|---|---|---|---|
| Fe(1) | ½ | 0 | 0 | 1.0 |
| Fe(2) | 0 | ½ | 0 | 1.0 |
| Se | 0 | 0 | 0.756 | 1.0 |
| Te | ½ | ½ | 0.285 | 1.0 |

**Table 2: Results from the spin density functional calculations on $FeSe_{0.5}Te_{0.5}$.**

a| Total energy of each AFM state in reference to that of the paramagnetic (PM) state in units of meV/formula ($\Delta E$), magnetic moment of Fe in units of $\mu_B$ (M), and internal coordinates associated with anion height from the Fe layer ($z_{Se/Te}$) obtained by structural optimization.

| | $\Delta E$ | M | $z_{Se}$ | $z_{Te}$ |
|---|---|---|---|---|
| PM | 0 | - | -0.2242 (1.334 Å) | 0.2744 (1.633 Å) |
| G-type AFM | -147.54 | 2.14 | -0.2453 (1.460 Å) | 0.2906 (1.730 Å) |
| Stripe AFM | -198.61 | 2.30 | -0.2460 (1.474 Å) | 0.2925 (1.741Å) |
| Double-stripe AFM | -174.19 | 2.45 | -0.2564 (1.526 Å) | 0.3003 (1.78 Å) |

b| Displacements of the center of the Wannier function localized at Fe(1) and Fe(2) sites, where the z values from the Fe plane (z=0) are shown in the unit of a.u.. The largest shift is observed for the $d_{yz}$ orbital of Fe(1) and the $d_{zx}$ one of Fe(2) which are the ones that hybridized most strongly with the Te *p* orbitals.

| MLWF | <Fe(1)> (a.u.) z | <Fe(2)> (a. u.) |
|---|---|---|
| $d_{xy}$ | 0.057188529540 | 0.057190958975 |
| $d_{yz}$ | 0.179277684686 | 0.043166144631 |
| $d_{z^2}$ | 0.057672527487 | 0.057701642460 |
| $d_{zx}$ | 0.04315708064 | 0.179291217237 |
| $d_{x^2}$ | 0.092899952649 | 0.092953036438 |

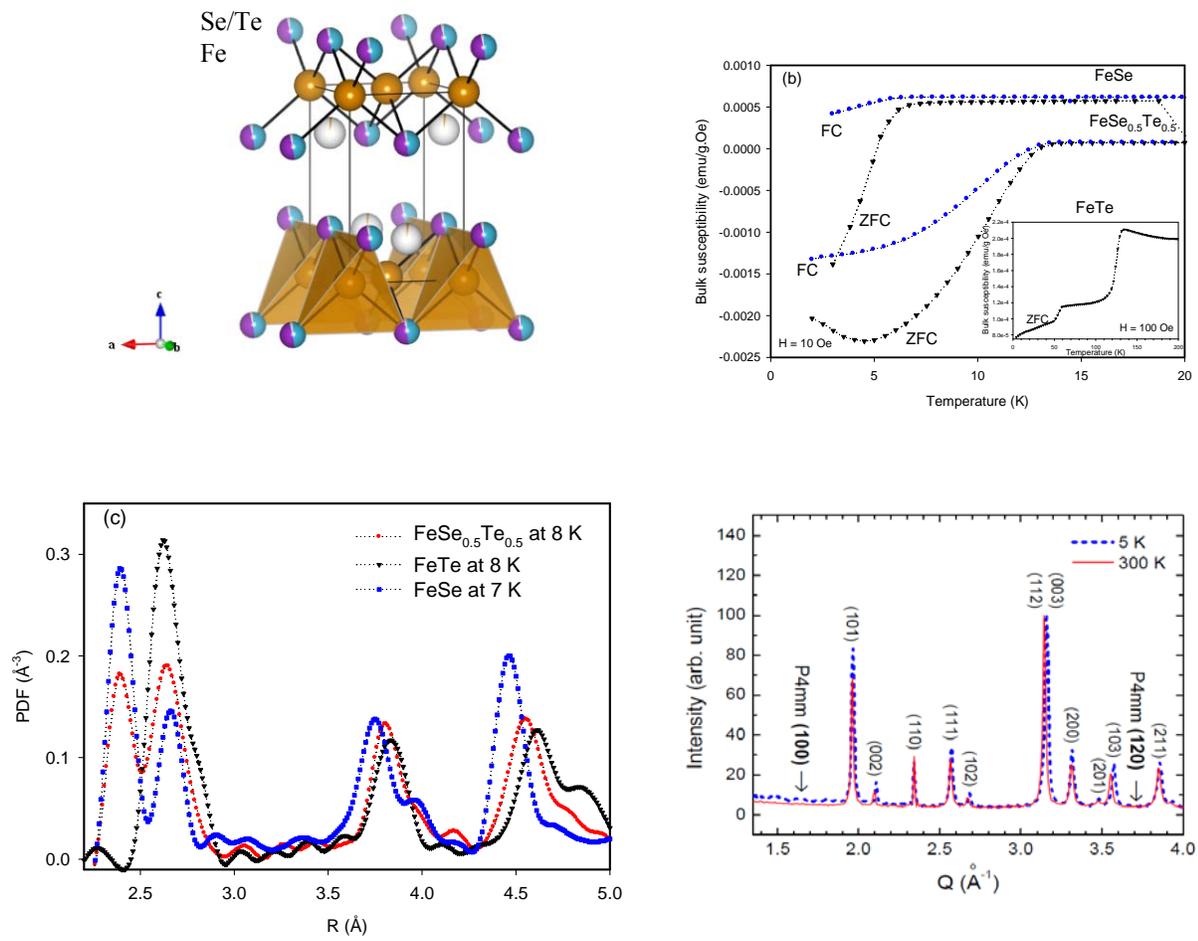

Fig. 1, Louca et al.

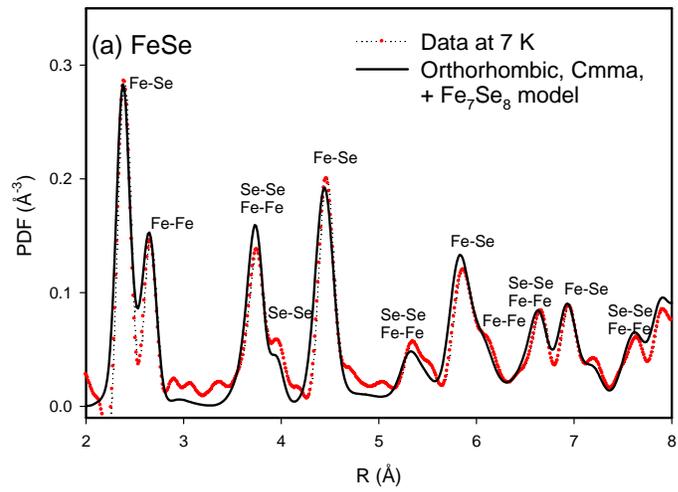

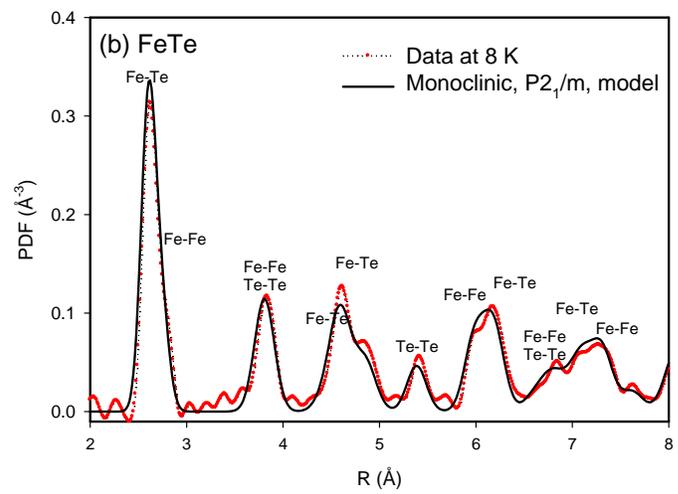

Fig. 2, Louca et al.

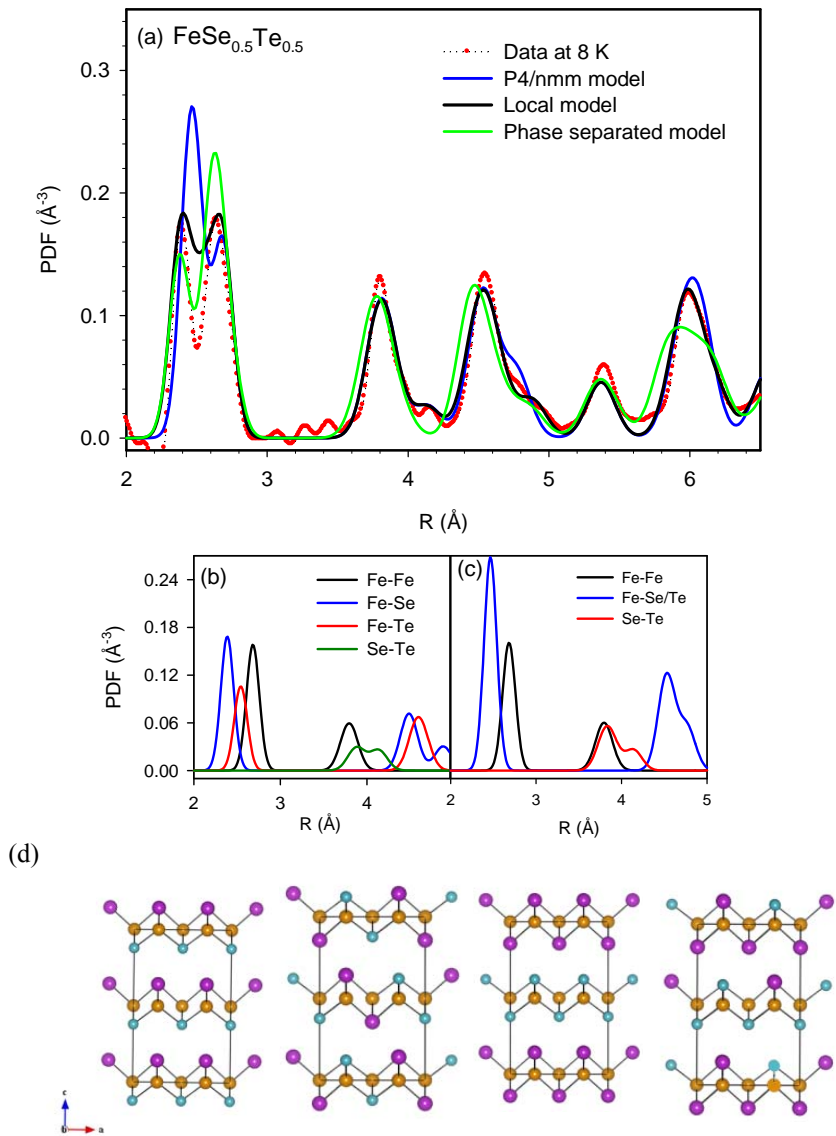

Fig. 3, Louca et al.